\input harvmac.tex

\def\text{}
\def\frac#1#2{ {#1 \over #2} }

\input epsf.tex

\lref\BerezhianiEWA{
  L.~Berezhiani and J.~Khoury,
[arXiv:1309.4461 [hep-th]].
}

\lref\WittenKN{
  E.~Witten,
[hep-th/0106109].
}
\lref\StromingerPN{
  A.~Strominger,
JHEP {\bf 0110}, 034 (2001).
[hep-th/0106113].
}

\lref\DeWittYK{
  B.~S.~DeWitt,
Phys.\ Rev.\  {\bf 160}, 1113 (1967)..
}

\lref\CreminelliYQ{
  P.~Creminelli and M.~Zaldarriaga,
JCAP {\bf 0410}, 006 (2004).
[astro-ph/0407059].
}

\lref\DeWittYK{
  B.~S.~DeWitt,
Phys.\ Rev.\  {\bf 160}, 1113 (1967)..
}

\lref\OsbornCR{
  H.~Osborn and A.~C.~Petkou,
Annals Phys.\  {\bf 231}, 311 (1994).
[hep-th/9307010].
}

\lref\MaldacenaVR{
  J.~M.~Maldacena,
JHEP {\bf 0305}, 013 (2003).
[astro-ph/0210603].
}

\lref\HartleAI{
  J.~B.~Hartle and S.~W.~Hawking,
Phys.\ Rev.\ D {\bf 28}, 2960 (1983)..
}

\lref\LarsenET{
  F.~Larsen, J.~P.~van der Schaar and R.~G.~Leigh,
JHEP {\bf 0204}, 047 (2002).
[hep-th/0202127].
}

\lref\MaldacenaNZ{
  J.~M.~Maldacena and G.~L.~Pimentel,
JHEP {\bf 1109}, 045 (2011).
[arXiv:1104.2846 [hep-th]].
}

\lref\HinterbichlerDPA{
  K.~Hinterbichler, L.~Hui and J.~Khoury,
[arXiv:1304.5527 [hep-th]].
}

\lref\deBoerXF{
  J.~de Boer, E.~P.~Verlinde and H.~L.~Verlinde,
JHEP {\bf 0008}, 003 (2000).
[hep-th/9912012].
}

\lref\SenatoreWY{
  L.~Senatore and M.~Zaldarriaga,
JCAP {\bf 1208}, 001 (2012).
[arXiv:1203.6884 [astro-ph.CO]].
}

\lref\AssassiZQ{
  V.~Assassi, D.~Baumann and D.~Green,
JCAP {\bf 1211}, 047 (2012).
[arXiv:1204.4207 [hep-th]].
}

\lref\LeblondYQ{
  L.~Leblond and E.~Pajer,
JCAP {\bf 1101}, 035 (2011).
[arXiv:1010.4565 [hep-th]].
}

\lref\CreminelliED{
  P.~Creminelli, J.~Norena and M.~Simonovic,
JCAP {\bf 1207}, 052 (2012).
[arXiv:1203.4595 [hep-th]].
}

\lref\PajerANA{
  E.~Pajer, F.~Schmidt and M.~Zaldarriaga,
[arXiv:1305.0824 [astro-ph.CO]].
}

\lref\CreminelliRH{
  P.~Creminelli, G.~D'Amico, M.~Musso and J.~Norena,
JCAP {\bf 1111}, 038 (2011).
[arXiv:1106.1462 [astro-ph.CO]].
}

\lref\WeinbergVY{
  S.~Weinberg,
Phys.\ Rev.\ D {\bf 72}, 043514 (2005).
[hep-th/0506236].
}

\lref\CheungSV{
  C.~Cheung, A.~L.~Fitzpatrick, J.~Kaplan and L.~Senatore,
JCAP {\bf 0802}, 021 (2008).
[arXiv:0709.0295 [hep-th]].
}

\lref\ChenNT{
  X.~Chen, M.~-x.~Huang, S.~Kachru and G.~Shiu,
JCAP {\bf 0701}, 002 (2007).
[hep-th/0605045].
}

\lref\GoldbergerRSA{
  W.~D.~Goldberger, L.~Hui and A.~Nicolis,
Phys.\ Rev.\ D {\bf 87}, 103520 (2013).
[arXiv:1303.1193 [hep-th]].
}

\lref\CreminelliCGA{
  P.~Creminelli, A.~Perko, L.~Senatore, M.~Simonovic and G.~Trevisan,
[arXiv:1307.0503 [astro-ph.CO]].
}

\lref\RouetUT{
  A.~Rouet and R.~Stora,
Lett.\ Nuovo Cim.\  {\bf 4S2}, 136 (1972), [Lett.\ Nuovo Cim.\  {\bf 4}, 136 (1972)]..
}

\lref\MansfieldPD{
  P.~Mansfield,
Nucl.\ Phys.\ B {\bf 418}, 113 (1994).
[hep-th/9308116].
}

\lref\SenatoreCF{
  L.~Senatore and M.~Zaldarriaga,
JHEP {\bf 1012}, 008 (2010).
[arXiv:0912.2734 [hep-th]].
}

\lref\SenatoreNQ{
  L.~Senatore and M.~Zaldarriaga,
JHEP {\bf 1301}, 109 (2013), [JHEP {\bf 1301}, 109 (2013)].
[arXiv:1203.6354 [hep-th]].
}

\lref\PimentelTW{
  G.~L.~Pimentel, L.~Senatore and M.~Zaldarriaga,
JHEP {\bf 1207}, 166 (2012).
[arXiv:1203.6651 [hep-th]].
}

\lref\SenatoreYA{
  L.~Senatore and M.~Zaldarriaga,
[arXiv:1210.6048 [hep-th]].
}

\lref\SenatoreWY{
  L.~Senatore and M.~Zaldarriaga,
JCAP {\bf 1208}, 001 (2012).
[arXiv:1203.6884 [astro-ph.CO]].
}

\lref\SalopekJQ{
  D.~S.~Salopek and J.~R.~Bond,
Phys.\ Rev.\ D {\bf 42}, 3936 (1990)..
}

\lref\SeeryAX{
  D.~Seery, M.~S.~Sloth and F.~Vernizzi,
JCAP {\bf 0903}, 018 (2009).
[arXiv:0811.3934 [astro-ph]].
}

\lref\LarsenPF{
  F.~Larsen and R.~McNees,
JHEP {\bf 0307}, 051 (2003).
[hep-th/0307026].
}

\lref\LarsenKF{
  F.~Larsen and R.~McNees,
JHEP {\bf 0407}, 062 (2004).
[hep-th/0402050].
}

\def\figin{\epsfcheck\figin}\def\figins{\epsfcheck\figins}
\def\epsfcheck{\ifx\epsfbox\UnDeFiNeD
\message{(NO epsf.tex, FIGURES WILL BE IGNORED)}
\gdef\figin##1{\vskip2in}\gdef\figins##1{\hskip.5in}
\else\message{(FIGURES WILL BE INCLUDED)}%
\gdef\figin##1{##1}\gdef\figins##1{##1}\fi}
\def\DefWarn#1{}
\def\figinsert{\goodbreak\midinsert}
\def\ifig#1#2#3{\DefWarn#1\xdef#1{fig.~\the\figno}
\writedef{#1\leftbracket fig.\noexpand~\the\figno}%
\figinsert\figin{\centerline{#3}}\medskip\centerline{\vbox{\baselineskip12pt
\advance\hsize by -1truein\noindent\footnotefont{\bf
Fig.~\the\figno:} #2}}
\bigskip\endinsert\global\advance\figno by1}

\rightline{PUPT-2452}
\Title{\vbox{\baselineskip12pt \hbox{} \hbox{
} }} {\vbox{\centerline{ Inflationary Consistency Conditions
  }
\centerline{from a Wavefunctional Perspective
  }
}}
\centerline{Guilherme L. Pimentel}
\bigskip

\centerline{ \it Joseph Henry Laboratories, Princeton University}
\centerline{\it Princeton, NJ 08544, USA}

\vskip .3in \noindent

It is shown that the squeezed limit of inflationary expectation values follows from reparametrization invariance of the wavefunction of the universe. This translates into a constraint on the longitudinal modes of functional derivatives of the wavefunction. Thus, the local non-Gaussianity induced by single field inflation is purely a gauge artifact. We focus on Einstein gravity in de Sitter space and single field inflation, although the formalism only relies on the diffeomorphism invariance of the theory, and thus applies to any theory of gravity.


 \Date{ }

\newsec{Introduction}

Recently, there has been much interest in calculating non-gaussian deviations for the statistics of primordial perturbations generated by inflation. Signatures of primordial non-gaussianity could falsify various models of the early universe. One is in general interested in computing three point expectation values of fields, evaluated at late times, when all modes have exited the horizon.

Maldacena pointed out \MaldacenaVR\ that there is a nice consistency check for the three point function of (single field) inflation. Namely, when one considers a ``squeezed" triangle shape, where one of the momenta is much smaller than the others (their sum needs to be zero due to translational invariance), the three point function can be written in terms of the tilt of the spectrum of the two point function. 

The intuition behind this consistency condition is as follows. In the squeezed regime, the long wavelength mode has exited the horizon earlier than the other modes, so its effect is to rescale the coordinates at which one computes the power spectrum for the other, shorter wavelength fluctuations. This intuition turns out to be correct for all models with a single field setting the natural ``clock" for the inflationary period \CreminelliYQ , and is thus a way to falsify single field inflation.

The consistency condition has been checked in many different models, and was derived in various different ways. An incomplete list of references is \refs{\CreminelliYQ,\ChenNT,\CheungSV}. The original consistency condition concerns only the leading term on the momentum of the long wavelength mode. Considerable progress has been made since, by several groups, on studying subleading corrections to the squeezed limit. Also, consistency conditions coming from soft internal momenta were found, relevant to squeezed limits of expectation values with four or more legs \refs{\SeeryAX, \CreminelliRH , \CreminelliED, \LeblondYQ,\SenatoreWY,\AssassiZQ,\GoldbergerRSA}. The derivations attack the problem from various perspectives. They either explore the broken symmetries of the (quasi-de Sitter) background, or some residual diffeomorphism invariance of the metric that was not completely fixed. There are also approaches that take the long mode as a classical background perturbation over which the shorter modes evolve.

In this paper, a different derivation of these results is provided. The object of primary interest will be the wavefunction of the universe, $\Psi[h,\phi]$, which has information on the probability for spacetime to have a spacelike slice with a given 3-metric and additional field profiles (for single field inflation, we also specify the profile of the inflaton on the slice). In this formalism, the wavefunction is specified by the so-called Wheeler-DeWitt equation \DeWittYK . We will show that coordinate reparametrization invariance of the three slice, also known as the momentum constraint, has all the information on squeezed limits of inflationary expectation values. In other words, all known consistency conditions follow from a symmetry of the wavefunction of the universe, or a constraint on its form. The wavefunctional perspective was also used to derive consistency conditions in \GoldbergerRSA .

Our situation here is analogous to the following in a gauge theory. We can compute Feynman diagrams and find correlation functions for the gauge fields $A_\mu$. These correlation functions satisfy some transversality condition, which basically removes the unphysical longitudinal modes from physical observables, like scattering amplitudes, and preserve unitarity etc. These are the Ward identities satisfied by the correlation functions. In gauge theory, we know what the good, gauge invariant observables should be (for example, correlations of field strengths $F_{\mu\nu}$, or Wilson loops). In gravity, a good observable should be diffeomorphism invariant. 

When we compute the expectation values, there are still ``longitudinal modes", or unphysical information, in these functions. The consistency condition basically tells us that the leading and next to leading order terms in the squeezed limit are fixed by this pure gauge information. From the point of view of a ``metaobserver" that sees our universe from outside, these would be unphysical modes. Because we have to pick a frame to make observations in cosmology, we would measure a squeezed non-gaussianity. The point is that it is fixed basically by the field content of the inflationary theory, and not from the details of the field interactions etc. This effect was recently computed and discussed in \PajerANA , in the context of translating the inflationary expectation values to CMB fluctuations in the sky. 

In \HinterbichlerDPA , it was pointed out that, in fact there is an infinite number of such consistency conditions, and, at each order in the long mode Taylor series, there should be terms fixed by diffeomorphism invariance. With the wavefunctional approach, that becomes very clear, as the consistency conditions can be extracted from a power series expansion of an expression of the schematic form:
\eqn\sqexu{
k_L^i{\delta^n\Psi\over\delta h_{ij}(k_L) \delta h (k_1) \delta h (k_2) ...} =  - k_1^j {\delta^{n-1}\Psi\over\delta h (k_1+k_L) \delta h (k_2) ...} -k_2^j{\delta^{n-1}\Psi\over\delta h (k_1) \delta h (k_2+k_L) ...}-\cdots
}
Where we omit indices of the other metric insertions for simplicity. These functional derivatives can be mapped to tree level expectation values of the fluctuating fields (metric, inflaton etc.). We can expand \sqexu\ around $k_L=0$ and, at each order, it will provide a consistency condition. In fact, \sqexu\ totally determines the form of the derivative of the wavefunction to leading and subleading orders. From quadratic order on, we cannot fully constrain this functional derivative, and that is when the truly physical contributions to the squeezed limit appear \refs{\PajerANA, \CreminelliCGA}. 

One nice feature of this wavefunctional formalism is that it is easily extended to theories with more fluctuating fields. Also, tracking the consequences of other symmetries of the wavefunction, like some flavor symmetry of the scalar sector, seems to be straightforward in this language. 

The paper is structured as follows. In section 2, we review the Wheeler-DeWitt formalism, in particular for Einstein gravity in de Sitter space, and for single field inflation. In section 3, we briefly treat the Hamiltonian constraints and comment on their implications. In section 4, we write the consistency condition from the wavefunctional perspective. In section 5, we derive the consequences for expectation values of fields, focusing on three point functions. In section 6, we make a few observations on gauge/gravity duality. In section 7 we discuss our results, followed by a few appendices on technical details. In particular, appendix B shows a somewhat simple but still interesting extension of the consistency condition to a single field inflation model with an additional massless scalar.

\newsec{Gravity in the Schrodinger picture: the Wheeler-DeWitt equations}

In the Wheeler-DeWitt approach to perturbative quantum gravity, the object of interest is the wavefunction of the universe. It gives the probability that the spacetime has a spatial slice with given 3-metric and field profile. The equations express the time and space reparametrization invariance of the wavefunction. The spatial reparametrization invariance implies the so-called momentum constraint on the wavefunction, and is an expression analogous to Gauss' law in electromagnetism. The time reparametrization  invariance encodes the dynamics of the theory. These are properly described in the $3+1$ decomposition of the metric, or the ADM formalism.

In this section we review how to obtain the Wheeler-DeWitt equations from the ADM decomposition of the metric. We analyze two particular cases of interest, namely, Einstein gravity in de Sitter space and single field inflation. 
\subsec{Einstein Gravity with positive Cosmological constant}
Start from the action:
\eqn\clasga{
S=\kappa\int\sqrt{-g}({}^{(4)}R-2\Lambda)
}
With $\kappa\equiv(16\pi G_N)^{-1}=\frac{M_{Pl}^2}{2}$. Then, in the ADM decomposition, $ds^2=-N^2 dt^2+h_{ij}(N^i dt + dx^i)(N^j dt + dx^j)$, the action reads:
\eqn\clasadm{
S=\kappa \int N\sqrt{h}\left[K_{ij}K^{ij}-K^2+ R-2\Lambda\right],~~K_{ij}\equiv\frac{1}{2N}(\dot{h}_{ij}-\nabla_iN_j-\nabla_jN_i)
}
Indices in \clasadm\ and from now on are raised with $h^{ij}$ instead of the 4D metric. We are omitting some boundary contributions which were subtracted by the Gibbons-Hawking-York term. The conjugate momenta to the metric are:
\eqn\conjmom{\eqalign{
&\pi\equiv\frac{\delta L}{\delta \dot N}=0,~~~~\pi^i\equiv\frac{\delta L}{\delta \dot N_i}=0\cr
&\pi^{ij}\equiv\frac{\delta L}{\delta \dot h_{ij}}=\kappa\sqrt{h}(K^{ij}-h^{ij}K)
}}
So the Hamiltonian will be of the form:
\eqn\hampg{\eqalign{
&H=\int\left\{N\left[\frac{1}{2 \kappa \sqrt{h}} G_{ij,kl}\pi^{ij}\pi^{kl}-\kappa\sqrt{h}(R-2\Lambda)\right]+2\nabla_i N_j\pi^{ij} \right\}\cr
&G_{ij,kl}\equiv(h_{ik}h_{jl}+h_{il}h_{jk}-h_{ij}h_{kl})
}}
$G_{ij,kl}$ is called the DeWitt metric\foot{In the literature, the factor of $1/\sqrt{h}$ is usually absorbed in the definition of $G$, so $G_{here}=\sqrt{h}G_{DeWitt}$}. Quantization on a basis that is diagonal in the three metric $h_{ij}$ corresponds to promoting $\pi^{ij}\to -i\hbar\frac{\delta}{\delta h_{ij}}$. Then, variation with respect to the lapse and shift yields the equations:
\eqn\WdWpgo{\eqalign{
&\left[\frac{\hbar^2}{2 \kappa \sqrt{h}}G_{ij,kl}{\delta\over \delta h_{ij}}{\delta \over \delta h_{kl}}+\kappa \sqrt{h}(R-2\Lambda)\right]\Psi=0 \cr
&-2 i \hbar \nabla_i\left[\frac{1}{\sqrt{h}}\frac{\delta \Psi}{\delta h_{ij}} \right]=0
} }
\subsec{Gravity plus a Scalar field}
Now write the action as follows:
\eqn\classf{
S=\kappa \int(R-(\nabla\phi)^2-2V(\phi))
}
Again, using the ADM decomposition, the action reads:
\eqn\clasfadm{
S=\kappa \int \sqrt{h}\left[N(K_{ij}K^{ij}-K^2+ R)+{1\over N}(\dot\phi-N^i\partial_i\phi)^2-Nh^{ij}\partial_i\phi\partial_j\phi-2NV(\phi) \right]
}
The conjugate momentum for the metric is the same, and the gravitational Hamiltonian is the same, but for the cosmological constant. The conjugate momentum for the scalar field is:
\eqn\momsc{
\pi_\phi={2\kappa\sqrt{h}\over N}(\dot\phi-N^i\partial_i\phi)
}
The total Hamiltonian will be:
\eqn\hamsc{\eqalign{
H=\int&\left\{N\left[\frac{1}{2 \kappa \sqrt{h}}G_{ij,kl}\pi^{ij}\pi^{kl}-\kappa\sqrt{h}R +\frac{1}{4\kappa\sqrt{h}}\pi_\phi^2+\kappa\sqrt{h}\left(h^{ij}\partial_i\phi\partial_j\phi+2V(\phi)\right)\right]+\right.\cr &\left.+2\nabla_i N_j\pi^{ij}+h^{ij}N_j\partial_i\phi\pi_\phi \right\}
}}
Now, the wavefunction $\Psi[h,\phi]$ is subject to the constraints:
\eqn\WdWsc{\eqalign{
&\left[\frac{\hbar^2}{2 \kappa \sqrt{h}}G_{ij,kl}{\delta\over \delta h_{ij}}{\delta \over \delta h_{kl}}+\kappa\sqrt{h}R +\frac{\hbar^2}{4\kappa\sqrt{h}}{\delta^2\over\delta\phi^2}-\kappa\sqrt{h}\left(h^{ij}\partial_i\phi\partial_j\phi+2V(\phi)\right)\right]\Psi=0\cr
&-2 i \hbar \nabla_i\left[\frac{1}{\sqrt{h}}\frac{\delta \Psi}{\delta h_{ij}} \right]+i\hbar{1\over\sqrt h}h^{ij}\partial_i\phi\frac{\delta \Psi}{\delta \phi}=0
}
}

\subsec{Tree level Wheeler-DeWitt equation}
Write $\Psi={\rm Exp}\{i W/\hbar \}$ and expand the equations \WdWpgo\ \WdWsc\ to zeroth order in $\hbar$. We get Hamilton-Jacobi equations for $W$, of the form:
\eqn\WdWpgtr{
\eqalign{
&\left[-\frac{1}{2 \kappa \sqrt{h}}G_{ij,kl}{\delta W\over \delta h_{ij}}{\delta W\over \delta h_{kl}}+\kappa \sqrt{h}(R-2\Lambda)\right]=0 \cr
&2 \nabla_i\left[\frac{1}{\sqrt{h}}\frac{\delta W}{\delta h_{ij}} \right]=0
} 
}
And, for single field inflation, of the form:
\eqn\WdWsctr{\eqalign{
&\left[-\frac{1}{2 \kappa \sqrt{h}}G_{ij,kl}{\delta W\over \delta h_{ij}}{\delta W\over \delta h_{kl}}+\kappa\sqrt{h}R -\frac{1}{4\kappa\sqrt{h}}\left({\delta W\over\delta\phi}\right)^2-\kappa\sqrt{h}\left(h^{ij}\partial_i\phi\partial_j\phi+2V(\phi)\right)\right]=0\cr
&2 \nabla_i\left[\frac{1}{\sqrt{h}}\frac{\delta W}{\delta h_{ij}} \right]-{1\over\sqrt h}h^{ij}\partial_i\phi\frac{\delta W}{\delta \phi}=0
}
}

\newsec{Structure of the Wavefunction at large ``volume" and time independence}
We are interested in computing the wavefunction at late times. Time is absent in the Wheeler-DeWitt approach to quantum gravity, so, in the context of inflation, we will be looking at the wavefunction for a spatial slice with large ``volume". In other words, take the spatial metric and redefine it as $h_{ij}= a^2 \hat {h}_{ij}$. We can then consider the asymptotics of \WdWpgtr\ and \WdWsctr\ as $a \to \infty$.

The time reparametrization constraint of general relativity is encoded in the Hamiltonian constraint. In principle, it will fix the wavefunction, given suitable boundary conditions. Here we just want to point out that there is a ``time-independent" piece of the wavefunction, which is nonlocal and encodes the superhorizon fluctuations in inflation.

\subsec{Pure gravity}
Begin by substituting $h_{ij}= a^2 \hat{h}_{ij}$ to \WdWpgtr . The Hamiltonian constraint becomes:
\eqn\scal{
\left[-\frac{1}{2 \kappa a^3 \sqrt{\hat h}}\hat{G}_{ij,kl}{\delta W\over \delta \hat h_{ij}}{\delta W\over \delta \hat h_{kl}}+\kappa \sqrt{\hat h}(a \hat R-a^3(2\Lambda))\right]=0
}

Now write $W=a^3 \alpha \int \sqrt{\hat h} + a \beta \int \sqrt{\hat h} \hat R + W_0+{\cal O}(1/a)$. Solving \scal\ order by order in $a$, we get:
\eqn\wloc{\eqalign{
&\alpha=4\kappa \sqrt{\frac{\Lambda}{3}}\cr&\beta=-\kappa\sqrt{\frac{3}{\Lambda}}\cr
&\hat h_{ij}\frac{\delta W_0}{\delta \hat h_{ij}}=0
}}
So \wloc\ tells us that $W_0$ is scale invariant, or $a$ independent. Note that, as we consider the $a \to \infty$ limit and compute expectation values, only $W_0$ is important, as the local terms are imaginary, so they will not appear in $|\Psi[h]|^2={\rm Exp}[2{\rm Re}\left({i\over \hbar} W_0\right)]$.

\subsec{Single Field Inflation}

Here the method is essentially the same, though the structure of the wavefunction is more complicated. A similar construction was carried for an arbitrary number of scalars, for a 5-D spacetime, in \deBoerXF . See also \refs{\LarsenPF , \LarsenKF} for a detailed analysis of the Hamilton-Jacobi equation for single field inflation. The Hamiltonian constraint is:
\eqn\hamsfsc{\eqalign{
&\left[-\frac{1}{2 \kappa a^3 \sqrt{h}}\hat G_{ij,kl}{\delta W\over \delta \hat h_{ij}}{\delta W\over \delta \hat h_{kl}}+\kappa\sqrt{\hat h} a R -\frac{1}{4\kappa a^3 \sqrt{\hat h}}\left({\delta W\over\delta\phi}\right)^2-\kappa\sqrt{\hat h}\left(a \hat h^{ij}\partial_i\phi\partial_j\phi+2a ^3 V(\phi)\right)\right]=0\cr
}
}

Now write $W=a^3 \int \sqrt{\hat h} U(\phi) + a \int \sqrt{\hat h} \left[\Phi(\phi)\hat R +\Theta(\phi) (h^{ij}\partial_i\phi\partial_j\phi)\right]+ W_0+{\cal O}(1/a)$. Solving \hamsfsc\ order by order in $a$, we get:
\eqn\wlocsc{\eqalign{
&V=\frac{1}{8\kappa^2}\left[\frac{3 U^2}{2}-U'^2 \right]\cr
&\frac{U\Phi}{2}-U'\Phi'=-2\kappa^2\cr
&\frac{U'\Theta'}{2}-U\left(\Phi''-\frac{\Theta}{4}\right)=\kappa^2\cr
&\frac{U'}{U}=\frac{\Phi'}{\Theta}\cr
&\hat h_{ij}\frac{\delta W_0}{\delta \hat h_{ij}}=\frac{U'}{U}\frac{\delta W_0}{\delta \phi}
}}
The auxiliary potential $U$ is related to the potential $V$ via \wlocsc. The variation of $W_0$ with respect to the trace of the metric is related to a variation of $W_0$ with respect to the scalar field. This relates two different gauge choices, one in which the trace of the metric is a fluctuating degree of freedom ($\zeta$ gauge) and the other where the scalar field fluctuates ($\delta\phi$ gauge). The factor that relates one to the other is related to the slow-roll parameter of single field inflation\refs{\LarsenET , \MaldacenaVR , \LarsenKF}.

The existence of a ``time-independent" piece of the wavefunction, for large volume (late times), is equivalent to the statement that there are fluctuations of the metric that freeze at late times \SalopekJQ . Those are the fluctuating fields whose correlations are calculated using the in-in formalism in inflation.

With Hartle-Hawking boundary conditions, $W_0$ can be computed explicitly. One evaluates the classical action with a solution for the equations of motion that obeys these boundary conditions. At very early times, the modes are in their flat space vacuum, as their physical wavelength is too small to probe any curvature effects of spacetime. $W_0$ has an imaginary piece that gives the tree level contribution to inflationary expectation values \refs{\MaldacenaVR,\HartleAI}.
\newsec{Consistency condition as a Ward Identity for derivatives of the Wavefunction}

In the previous section, we showed that the wavefunction has a piece that is late time independent. Now we want to show that the consistency condition for the cosmological correlators arises from reparametrization invariance of the wavefunction of the universe, in particular, of $W_0$. We write $h_{ij}=a^2(\delta_{ij}+p_{ij})$ and consider the limit of $a\to \infty$, which would correspond to a late time slice in the semiclassical approximation.

Now we impose that the wavefunction is invariant under spatial diffeomorphisms. This means that $\Psi[h_{ij}+\nabla_{(i}v_{j)}]=\Psi[h_{ij}]$. This implies that:

\eqn\wig{
\delta \Psi [h_{ij}] = 2 \int d^3x \nabla_a v_b(x) \frac{\delta \Psi[h_{ij}]}{\delta h_{ab}(x)}=0 \Rightarrow \nabla_{i}\left[\frac{1}{\sqrt{h(x)}} \frac{\delta \Psi}{\delta h_{ij}(x)}=0  \right]
}

Equation \wig\ is the Ward identity for the wavefunction of the universe. It is a statement on its reparametrization invariance. Of course, this is the same as equation \WdWpgo . Specializing to $ W_0$, the scale invariant piece of the wavefunction, as in \wloc , we see that it also satisfies \wig\ with $\Psi \to  W_0$, as the other terms that survive in the $a\to \infty$ limit automatically satisfy \wig , as they are invariant under spatial reparametrizations.

Let us now perform a perturbative expansion in $ W_0$. We take the perturbations to be around the flat metric, as $\delta_{ij}+p_{ij}$. Of course, this is due to what we know about the universe being approximately flat after inflation. The consistency conditions can be easily generalized to expansions around different backgrounds, as the WdW equations are invariant under the choice of background metric.

As we are interested in the non-local piece of the wavefunction, we start quadratic in the metric\foot{In principle there can be a scale invariant, local contribution to the wavefunction, proportional to the gravitational Chern-Simons term. It can be argued that this term is a pure phase \MaldacenaNZ, and thus will not contribute to the sort of correlators we consider here. In any case, even if it were real, it would contribute a local term to expectation values, and we are interested in consistency conditions for the nonlocal terms.}:
\eqn\pertexp{\eqalign{
& W_0[\hat h]= \frac{1}{2!}  \int d^3x d^3y \left(\frac{ \delta^2  W_0[\delta]}{\delta \hat h_{ab}(x)\delta \hat h_{cd}(y)}\right) p_{ab}(x) p_{cd}(y)+ \cr &+ \frac{1}{3!} \int d^3x d^3y d^3z \left(\frac{ \delta^3  W_0[\delta]}{\delta \hat h_{ab}(x)\delta \hat h_{cd}(y)\hat h_{ef}(z)}\right)  p_{ab}(x) p_{cd}(y)p_{ef}(z)+\cr& + \frac{1}{4!} \int d^3x d^3y d^3z d^3w \frac{ \delta^4  W_0[\delta]}{\delta \hat h_{ab}(x)\delta \hat h_{cd}(y)\delta\hat h_{ef}(z)\delta\hat h_{ij}(w)} p_{ab}(x) p_{cd}(y)p_{ef}(z)p_{ij}(w)+\cdots
}}
We are Taylor expanding around the flat metric. Square brackets in the derivatives mean that we calculate the derivative at the background metric. As we will be mostly dealing with $W_0$ from here on, we will omit the hat on $\hat h_{ij}$.

We now work out the consequences of \wig\ to the coefficients in the perturbative expansion \pertexp . The idea is to commute an insertion of $\delta/\delta \hat h_{ij}$ into \wig\ and then evaluate the resultant expression for $h_{ij}=\delta_{ij}$. We rewrite \wig\ as:

\eqn\wigex{
\nabla_i\left[\frac{\delta W_0}{\delta{ h_{ij}(x)}}\right]=\partial_i\left(\frac{\delta W_0}{\delta{ h_{ij}(x)}}\right)+\Gamma^j_{ik}(x)\frac{\delta W_0}{\delta{h_{ik}(x)}}=0
}
The only issue here is how to commute through the Christoffel symbol. Writing $\Gamma^a_{bc}= h^{ad}\Gamma_{dbc}$ the following expressions are useful:
\eqn\chrvar{\eqalign{
&\frac{\delta^n\Gamma^a_{bc}(x)}{\delta h_{i_1j_1}(y_1)\cdots\delta h_{i_nj_n}(y_n)}=\frac{\delta^n h^{ad}(x)}{\delta h_{i_1j_1}(y_1)\cdots\delta h_{i_nj_n}(y_n)}\Gamma_{dbc}(x)+\cdots\cr &\cdots+\sum_{k=1}^n\frac{\delta^{n-1}h^{ad}(x)}{\delta h_{i_1j_1}(y_1)\cdots\delta h_{i_nj_n}(y_n)}\frac{\delta \Gamma_{dbc}(x)}{\delta h_{i_k j_k}(y_k)}\cr
&\frac{\delta h^{ab}(x)}{\delta h_{cd}(y)}=-h^{am}(x)h^{bn}(x)\delta^{cd}_{mn}\delta(x-y),~~\delta^{cd}_{mn}\equiv\frac{1}{2}(\delta^c_m\delta^d_n+\delta^c_n\delta^d_m)\cr
&\frac{\delta\Gamma_{dbc}(x)}{\delta h_{ij}(y)}=\frac{1}{2}\left(\delta^{ij}_{bd}~\partial^x_c\delta(x-y)+\delta^{ij}_{cd}~\partial^x_b\delta(x-y)-\delta^{ij}_{bc}~\partial^x_d\delta(x-y)\right)
}}

{\it Second derivative}

First let us just commute one insertion of $\delta/\delta h_{ij}$ through \wig . We get:

\eqn\trans{
\frac{\partial}{\partial x^i} \left(\frac{\delta^2 W_0[\delta]}{\delta h_{ij}(x)\delta h_{kl}(y)} \right)=0
}

{\it Third derivative}

We now commute two insertions of $\delta/\delta h_{ij}$ through \wig . We get:
\eqn\trant{\eqalign{
&\frac{\partial}{\partial x^i} \left(\frac{\delta^3 W_0[\delta]}{\delta h_{ij}(x)\delta h_{kl}(y)\delta h_{mn}(z)} \right)=-\frac{1}{2}\left[\left(\delta^{jk}\frac{\delta^2 W_0[\delta]}{\delta h_{il}(x)\delta h_{mn}(z)}\frac{\partial}{\partial x^i}\delta(x-y) +\{k\leftrightarrow l\}\right)+\right.\cr&+\left.\left(\delta^{jm}\frac{\delta^2 W_0[\delta]}{\delta h_{kl}(y)\delta h_{in}(x)}\frac{\partial}{\partial x^i}\delta(x-z) + \{m\leftrightarrow n\}\right)-\left(\frac{\delta^2 W_0[\delta]}{\delta h_{kl}(x)\delta h_{mn}(z)}\frac{\partial}{\partial x_j}\delta(x-y)+\right.\right.\cr&\left.\left.+\frac{\delta^2 W_0[\delta]}{\delta h_{kl}(y)\delta h_{mn}(x)}\frac{\partial}{\partial x_j}\delta(x-z)\right)\right]
}}

\newsec{Consequences for expectation values}

The expectation values of insertions of the metric is given by:
\eqn\expv{
\langle  h_{ij}(x) h_{kl}(y) \cdots \rangle = \int dh |\Psi[h]|^2 h_{ij}(x) h_{kl}(y) \cdots 
}

It is convenient to do the following before moving on. We want to write the expectation values of operators in momentum space. We also use a basis of polarization tensors that is traceless and transverse with respect to the flat metric:
\eqn\poltens{
\epsilon^s_{ij}\epsilon^{s'}_{ij}=2\delta^{s s'},~~~~~k_i\epsilon^s_{ij}=0
}
Indices are contracted with $\delta^{ij}$. We call the helicity modes $+$ and $-$.
Angular momentum conservation and momentum conservation tells us that the only two point functions allowed are $\langle p^+ p^+\rangle$ and $\langle p^- p^-\rangle$, with the perturbation being $p_{ij}\equiv h_{ij}-\delta_{ij}$. Now write the wavefunction as follows:
\eqn\quantwf{
\Psi=\Psi_{local}\times{\rm Exp}\left\{\sum_n \int d k_1 \cdots dk_n \frac{1}{n!}\frac{\delta^n W_0[\delta]}{\delta h^{s_1}(k_1)\cdots \delta h^{s_n}(k_n)} p^{s_1}(k_1)\cdots p^{s_n}(k_n) \right\}
}

In terms of \quantwf , the two point expectation value for gravitational wave perturbations is given by:

\eqn\tpf{
\langle p^{s_1}(k_1) p^{s_2}(k_2) \rangle=-\frac{1}{2 {\rm Re}\frac{\delta^2 W_0[\delta]}{\delta h^{s_1}(k_1) \delta h^{s_2}(k_2)}}
}

{\it Three Point Function}

In terms of \quantwf , the three point expectation value for gravitational wave perturbations is given by:

\eqn\tpf{
\langle p^{s_1}(k_1) p^{s_2}(k_2) p^{s_3}(k_3)\rangle=-\frac{2 {\rm Re} \left(\frac{\delta^3 W_0[\delta]}{\delta h^{s_1}(k_1) \delta h^{s_2}(k_2)\delta h^{s_3}(k_3)}\right)}{\Pi_i {\rm Re}\left(2\frac{\delta^2 W_0[\delta]}{\delta h^{s_i}(k_i) \delta h^{s_i}(-k_i)}\right)}
}

Let us now understand how \trant\ constrains the shape of the three point function in the squeezed limit. Start from looking at \trans\ and \trant\ in momentum space. As the background is translation invariant, the derivatives of the wavefunction should only depend on distances between points. In momentum space, this means that there is always a subtended momentum conserving delta function in front of an expectation value\foot{The notation used in \MaldacenaVR\ and in many papers in the literature is to use a $'$ in front of the expectation value, e. g.  $\langle h(k_1) h(k_2)\rangle = \delta(k_1+k_2) \langle h(k_1) h(-k_1) \rangle '$. We will always assume that the delta function is taken care of, and use it to eliminate one momentum variable in the various expectation values.}. Thus, an $n$-point expectation value will explicitly depend on $n-1$ momenta, the last momentum dependence removed by translation invariance. 

Before giving the final forms for \trans\ and \trant , we need to do one more thing. The variable we actually use in the bulk computations is $\gamma_{ij}$, such that $h_{ij}={\rm Exp}(\gamma)_{ij}$. To cubic order, $h_{ij}=\delta_{ij}+\gamma_{ij}+\frac{1}{2}\gamma_{ik}\gamma_{kj}$. Translating the relation for the three point derivative \trant\ will induce new contact terms in the Ward identity, due to the use of the chain rule. In momentum space, \trans\ and \trant\ will read:
\eqn\transmom{
k_{1,i}\frac{\delta^2 W_0[\delta]}{\delta \gamma_{ij}(k_1)\delta \gamma_{kl}(k_2)}=0
}
\eqn\trantmom{\eqalign{
k_{1,a}&\frac{\delta^3 W_0[\delta]}{\delta \gamma_{aj}(k_1)\delta \gamma_{kl}(k_2)\delta \gamma_{mn}(k_3)}=\cr&=\frac{1}{2}\left[\delta^{jk}k_{2,a}\frac{\delta^2 W_0[\delta]}{\delta \gamma_{al}(-k_3)\delta \gamma_{mn}(k_3)} + \delta^{jl} k_{2,a} \frac{\delta^2 W_0[\delta]}{\delta \gamma_{ak}(-k_3)\delta \gamma_{mn}(k_3)} + \right.\cr &\left.+ \delta^{jm}k_{3,a}\frac{\delta^2 W_0[\delta]}{\delta \gamma_{kl}(k_2)\delta \gamma_{an}(-k_2)} + \delta^{jn} k_{3,a} \frac{\delta^2 W_0[\delta]}{\delta \gamma_{kl}(k_2)\delta \gamma_{ma}(-k_2)}-\right. \cr &\left. - k_{2,j}\frac{\delta^2 W_0[\delta]}{\delta \gamma_{kl}(-k_3)\delta \gamma_{mn}(k_3)} -k_{3,j}\frac{\delta^2 W_0[\delta]}{\delta \gamma_{kl}(k_2)\delta \gamma_{mn}(-k_2)}\right]+\cr &+k_{1,a}\left[\delta^{kl}_{bd}\delta^{aj}_{dc} \frac{\delta^2 W_0[\delta]}{\delta\gamma_{bc}(-k_3)\delta\gamma_{mn}(k_3)}+\delta^{mn}_{bd}\delta^{ij}_{dc} \frac{\delta^2 W_0[\delta]}{\delta\gamma_{kl}(k_2)\delta\gamma_{bc}(-k_2)} \right]
}}
The last line of \trantmom\ comes from the change of variables $p\to\gamma$. The derivatives are evaluated around the flat background, meaning that we set $\gamma=0$ after taking the derivative. Dummy indices are $a,\cdots, d$ in \trantmom . \trantmom\ encodes all consistency conditions for the three point function of inflationary perturbations.

\subsec{Extracting the consistency condition}
To get the consistency conditions of inflation, we need to consider the squeezed limit, or $k_1 \to 0$. We want to show that \trantmom\ implies an infinite number of such consistency conditions, as recently discussed by \HinterbichlerDPA . The leading correction to the consistency condition, which is also completely fixed by the longitudinal modes, was first discussed in \CreminelliED . Now, all one needs to do is to Taylor expand  \trantmom\ around $k_1=0$. Let us introduce the simplified notation:

\eqn\notch{
D^{ij}_k W_0\equiv {\delta  W_0[\delta]\over\delta \gamma_{ij}(k)}
}

Then, the three point function for gravitational waves is given by:
\eqn\tpfg{
\langle \gamma^{s_1}_{k_1}\gamma^{s_2}_{k_2}\gamma^{s_3}_{k_3}\rangle = -\frac{2 {\rm Re} \left[D^{s_1}_{k_1} D^{s_2}_{k_2} D^{s_3}_{k_3}  W_0\right]}{2 {\rm Re} \left[D^{s_1}_{k_1}D^{s_1}_{-k_1} W_0\right] 2 {\rm Re} \left[D^{s_2}_{k_2}D^{s_2}_{-k_2} W_0\right] 2 {\rm Re} \left[D^{s_3}_{k_3}D^{s_3}_{-k_3} W_0\right] }
}
And the consistency condition is encoded in the identity:
\eqn\conall{\eqalign{
k_1^a&D^{aj}_{k_1}D^{kl}_{k_2}D^{mn}_{k_3} W_0=\cr&=\frac{1}{2}\left[\delta^{jk}k_2^a D^{al}_{-k_3} D^{mn}_{k_3} W_0 + \delta^{jl} k_2^a D^{ak}_{-k_3}D^{mn}_{k_3} W_0 +  \delta^{jm}k^a_3 D^{kl}_{k_2}D^{an}_{-k_2} W_0 + \delta^{jn} k^a_3 D^{kl}_{k_2}D^{ma}_{-k_2} W_0-\right. \cr &\left. - k^j_2 D^{kl}_{-k_3}D^{mn}_{k_3} W_0 -k^j_3D^{kl}_{k_2}D^{mn}_{-k_2} W_0\right]+k^a_1\left[\delta^{kl}_{bd}\delta^{aj}_{dc} D^{bc}_{-k_3}D^{mn}_{k_3} W_0+\delta^{mn}_{bd}\delta^{aj}_{dc} D^{kl}_{k_2}D^{bc}_{-k_2} W_0 \right] 
}}

Now we expand \tpfg\ and \conall\ around $k_1=0$. Assuming that $\lim_{k_1\to 0} k_1^aD^{aj}_{k_1}D^{kl}_{k_2}D^{mn}_{k_3} W_0=0$, which is true if there are no terms of the form $\log k_1$ in the three point function of cosmological perturbations, we get:
\eqn\conallfo{\eqalign{
&\lim_{k_1\to0}D^{ij}_{k_1}D^{kl}_{k_2}D^{mn}_{-k_1-k_2} W_0=\cr&=-\frac{1}{2}\left[\delta^{jk}D^{il}_{-k_2} D^{mn}_{k_2} W_0+\delta^{jl}D^{ik}_{-k_2} D^{mn}_{k_2} W_0+\delta^{jm}D^{kl}_{-k_2} D^{in}_{k_2} W_0+\delta^{jn}D^{kl}_{-k_2} D^{mi}_{k_2} W_0+\right. \cr &\left.+k_{2,j}\frac{\partial}{\partial k_{2,i}}D^{kl}_{-k_2} D^{mn}_{k_2} W_0-\delta^{ij}D^{kl}_{-k_2} D^{mn}_{k_2} W_0\right]+\left[\delta^{kl}_{bd}\delta^{ij}_{cd}D^{bc}_{-k_2} D^{mn}_{k_2} W_0+\delta^{mn}_{bd}\delta^{ij}_{cd}D^{kl}_{-k_2} D^{bc}_{k_2} W_0\right]
}}
Now, we contract \conallfo\ with polarization tensors for the fluctuations. We obtain:
\eqn\conallfopt{
\lim_{k_1\to0}D^{s_1}_{k_1}D^{s_2}_{k_2}D^{s_2}_{-k_1-k_2} W_0=-{1\over2}\epsilon^{ij}_{1}k^i_{2}{\partial\over\partial k^j_{2}}D^{s_2}_{-k_2} D^{s_2}_{k_2} W_0=-\epsilon^{ij}_{1}k_{2}^ik_{2}^j{\partial\over\partial k^2_{2}}D^{s_2}_{-k_2} D^{s_2}_{k_2} W_0
}
We now see that the leading order contribution to the squeezed limit of the three point function of gravitational waves is:
\eqn\tpfsq{\eqalign{
&\lim_{k_1\to0}\langle \gamma^{s_1}_{k_1}\gamma^{s_2}_{k_2}\gamma^{s_3}_{k_3}\rangle = -\frac{2 {\rm Re} \lim_{k_1\to0}\left[D^{s_1}_{k_1} D^{s_2}_{k_2} D^{s_3}_{k_3}  W_0\right]\delta^{s_2s_3}}{2 {\rm Re} \left[D^{s_1}_{k_1}D^{s_1}_{-k_1} W_0\right] \left(2 {\rm Re} \left[D^{s_2}_{k_2}D^{s_2}_{-k_2} W_0\right]\right)^2  }=\cr&=-\left[-{1\over 2 {\rm Re}[D^{s_1}_{k_1}D^{s_1}_{-k_1} W_0]}\right]\left[-\epsilon^{ij}_1k_2^ik_2^j {\partial \over \partial k_2^2}\left({1\over 2 {\rm Re}[D^{s_2}_{k_2}D^{s_2}_{-k_2} W_0]}\right) \right]\delta^{s_2s_3}=\cr&=-\langle \gamma^{s_1}_{k_1}\gamma^{s_1}_{-k_1}\rangle \epsilon^{ij}_1k_2^ik_2^j {\partial \over \partial k_2^2}\langle \gamma^{s_2}_{k_2}\gamma^{s_2}_{-k_2}\rangle\delta^{s_2s_3}
}}
Which is the standard consistency condition for the gravitational wave three point function \MaldacenaVR .
\subsec{Scalar Fluctuations}
Although there is no scalar mode in pure gravity in de Sitter space, (as is illustrated by equation \wloc ) we can still make use of \conall\ to extract the consistency condition for the inflaton, $\zeta$. The reason is that we evaluate the wavefunction at a surface of constant background field, $ W_0[h,\phi]$ with $\partial_i\phi=0$. Then, the momentum constraint \WdWsc\ reduces to the one in pure gravity, and thus \conall\ applies. As the $\zeta$ mode is also taken as the exponential of the metric, but of its trace, instead of its traceless transverse component, all we need to do is to contract \conall\ with $\delta^{ij}$, etc. At the level of the three point function, we take our metric to be $h_{ij}=\delta_{ij}(1+2\zeta+2\zeta^2)$. That corresponds to the substitution $\gamma_{ij}\to 2\zeta \delta_{ij}$. In \conall\ that corresponds to $2 D^{ij}_k \delta^{ij}\to D_k$, where $D_k\equiv\delta/\delta \zeta_k$. We obtain:
\eqn\conallfoptsc{
\lim_{k_1\to0}D_{k_1}D_{k_2}D_{-k_1-k_2} W_0=\left[\left(3-k^i_{2}{\partial\over\partial k^i_{2}}\right)D_{-k_2} D_{k_2} W_0\right]=\left[\left(3-k_{2}{\partial\over\partial k_{2}}\right)D_{-k_2} D_{k_2} W_0\right]
}
Thus, for the three point function, we obtain:
\eqn\tpfsqsc{\eqalign{
&\lim_{k_1\to0}\langle \zeta_{k_1}\zeta_{k_2}\zeta_{k_3}\rangle = -\frac{2 {\rm Re} \lim_{k_1\to0}\left[D_{k_1} D_{k_2} D_{k_3}  W_0\right]}{2 {\rm Re} \left[D_{k_1}D_{-k_1} W_0\right] \left(2 {\rm Re} \left[D_{k_2}D_{-k_2} W_0\right]\right)^2  }=\cr&=-\left[-{1\over 2 {\rm Re}[D_{k_1}D_{-k_1} W_0]}\right]\left[-{1\over 2 {\rm Re}[D_{k_2}D_{-k_2} W_0]}\right]\left[-\frac{d\log\left(-2 {\rm Re} \left[D_{k_2}D_{-k_2} W_0 k_2^{-3}\right]\right)}{d \log k}\right] =\cr&=-\langle \zeta_{k_1}\zeta_{-k_1}\rangle \langle \zeta_{k_2}\zeta_{-k_2}\rangle {\partial \log \left[k_2^3  \langle \zeta_{k_2}\zeta_{-k_2}\rangle\right] \over \partial \log k_2}}}
\subsec{Mixed three point functions}
For three point functions of one long scalar and two short tensor fluctuations, and vice-versa, the derivation is essentially the same. One just needs to contract \conall\ with the proper polarization tensors etc. We just quote the final results. For a long scalar mode and short tensor modes we have:
\eqn\tpfmixlsc{\eqalign{
&\lim_{k_1\to0}\langle \zeta_{k_1}\gamma^{s_2}_{k_2}\gamma^{s_3}_{k_3}\rangle = -\frac{2 {\rm Re} \lim_{k_1\to0}\left[D_{k_1} D^{s_2}_{k_2} D^{s_3}_{k_3}  W_0\right]\delta^{s_2s_3}}{2 {\rm Re} \left[D_{k_1}D_{-k_1} W_0\right] \left(2 {\rm Re} \left[D^{s_2}_{k_2}D^{s_2}_{-k_2} W_0\right]\right)^2  }=\cr&=-\left[-{1\over 2 {\rm Re}[D_{k_1}D_{-k_1} W_0]}\right]\left[-{\delta^{s_2s_3}\over 2 {\rm Re}[D^{s_2}_{k_2}D^{s_2}_{-k_2} W_0]}\right]\left[-\frac{d\log\left(-2 {\rm Re} \left[D^{s_2}_{k_2}D^{s_2}_{-k_2} W_0 k_2^{-3}\right]\right)}{d \log k}\right] =\cr&=-\langle \zeta_{k_1}\zeta_{-k_1}\rangle \langle \gamma^{s_2}_{k_2}\gamma^{s_2}_{-k_2}\rangle\delta^{s_2s_3} {\partial \log \left[k_2^3  \langle \gamma^{s_2}_{k_2}\gamma^{s_2}_{-k_2}\rangle\right] \over \partial \log k_2}}}
While, for a long tensor mode and two short scalar modes, we get:
\eqn\tpfmixlt{\eqalign{
&\lim_{k_1\to0}\langle \gamma^{s_1}_{k_1}\zeta_{k_2}\zeta_{k_3}\rangle = -\frac{2 {\rm Re} \lim_{k_1\to0}\left[D^{s_1}_{k_1} D_{k_2} D_{k_3}  W_0\right]}{2 {\rm Re} \left[D^{s_1}_{k_1}D^{s_1}_{-k_1} W_0\right] \left(2 {\rm Re} \left[D_{k_2}D_{-k_2} W_0\right]\right)^2  }=\cr&=-\left[-{1\over 2 {\rm Re}[D^{s_1}_{k_1}D^{s_1}_{-k_1} W_0]}\right]\left[-\epsilon^{ij}_1k_2^ik_2^j {\partial \over \partial k_2^2}\left({1\over 2 {\rm Re}[D_{k_2}D_{-k_2} W_0]}\right) \right]=\cr&=-\langle \gamma^{s_1}_{k_1}\gamma^{s_1}_{-k_1}\rangle \epsilon^{ij}_1k_2^ik_2^j {\partial \over \partial k_2^2}\langle \zeta_{k_2}\zeta_{-k_2}\rangle}}

\subsec{Higher order consistency conditions}

Let us expand the three point function of fluctuations in a Taylor series around $k_1=0$. For simplicity, we consider scalar fluctuations:

\eqn\tayexp{
\langle \zeta_{k_1}\zeta_{k_2}\zeta_{-k_1-k_2}\rangle=  \langle\zeta_{k_1}\zeta_{-k_1}\rangle\left[Z(k_2)+k_1^aF^a(k_2)+{1\over2}k_1^ak_1^b S^{ab}(k_2)+\cdots\right]
}

It was already pointed out in \MaldacenaVR\ that the leading order term $Z(k_2)$ is fixed by the two point function, which is what we call the inflationary consistency condition.
In \CreminelliED\ it was argued that the first order term $F^a(k_2)$ is also completely fixed by some residual conformal symmetry of the background. Reference \HinterbichlerDPA\ studied constraints to the higher order terms in \tayexp , and found general Ward identities that should be obeyed by some combinations of gravitational wave and inflaton expectation values. 

All of these consistency conditions follow from a Taylor expansion of the longitudinal mode Ward identities \conall . So the inflationary consistency conditions can be explained by the reparametrization invariance, or momentum constraint, of the wavefunction of the universe. The terms that have physical content, and are probing the primordial non-gaussianity of inflationary perturbations, start quadratic in $k_1$ in $\tayexp$. In \PajerANA\ it was pointed out that the squeezed three point function of single field inflation gives rise to no effect in a physical observable. This is of course consistent with the picture that the squeezed limit is totally fixed by diffeomorphism invariance, as physical observables are diff-invariant. In other words, there is residual gauge symmetry in the squeezed limit of expectation values of inflationary fluctuations, and these can be tracked down from the original symmetry.

 Here we derive the consistency condition discussed in \CreminelliED , which completely fixes the linear term in $k_1$ in \tayexp . We also discuss the generalized consistency conditions of \HinterbichlerDPA , pointing out why from quadratic order on, the longitudinal modes do not fix completely the three point function. Note that our derivation makes no use of conformal symmetry; we rely purely on reparametrization invariance of the wavefunction.
 
First, contract \conall\ with $4\delta^{kl}\delta^{mn}$. We get\foot{In general, the Ward identity will look like $k_1^iD_{k_1}^{ij}D_{k_2}\cdots D_{k_n}W_0=-k_2^j D_{k_1+k_2}\cdots D_{k_n}W_0-\cdots-k_n^j D_{k_2}\cdots D_{k_n+k_1}W_0$, i.e., it relates the $n$-derivative to the $n-1$-derivative of the wavefunction, evaluated at shifted momenta \MaldacenaNZ .}:
\eqn\scgen{
k_1^aD^{aj}_{k_1}D_{k_2}D_{k_3} W_0={1\over2}\left[-k_2^j D_{k_3}D_{-k_3} W_0-k_3^jD_{k_2}D_{-k_2} W_0\right]
}
 
Now, take the first order correction to the three point function in the squeezed limit. We Taylor expand \tpfg , for scalar fluctuations, to first order in $k_1$. Comparing with the formula \tayexp\ we get:
\eqn\expfirst{\eqalign{
&Z(k_2)={1\over2}{{\rm Re}\lim_{k_1\to0} D_{k_1} D_{k_2}D_{k_3} W_0 \over({\rm Re} \left[D_{k_2}D_{-k_2} W_0\right])^2}=-\langle\zeta_{k_2}\zeta_{-k_2}\rangle{\partial \log \left[k_2^3  \langle \zeta_{k_2}\zeta_{-k_2}\rangle\right] \over \partial \log k_2}\cr
&F^a(k_2)={1\over2}\left({\lim_{k_1\to0}\partial_{k_1^a}{\rm Re} D_{k_1} D_{k_2}D_{k_3} W_0 \over({\rm Re} \left[D_{k_2}D_{-k_2} W_0\right])^2}-{{\rm Re}\lim_{k_1\to0} D_{k_1} D_{k_2}D_{k_3} W_0 \over({\rm Re} \left[D_{k_2}D_{-k_2} W_0\right])^3}\partial_{k_2^a}{\rm Re} \left[D_{k_2}D_{-k_2} W_0\right] \right)
}}

Now we take two derivatives of \conall\ with respect to $k_1$ and take $k_1\to0$. That will give:
\eqn\linconall{
{\partial\over\partial k_1^l}D^{ij}_{k_1}D_{k_2}D_{k_3} W_0+{\partial\over\partial k_1^i}D^{lj}_{k_1}D_{k_2}D_{k_3} W_0=-{1\over2}k_2^j{\partial^2\over\partial k_2^l\partial k_2^i}
D_{k_2}D_{-k_2} W_0}
Note that the index $j$ in \linconall\ is singled out, and the left hand side is symmetric in $i$, $l$. We contract \linconall\ with $2\delta^{ij}$ and $\delta^{il}$ and subtract the equations we obtain, getting:
\eqn\auxon{\eqalign{
\lim_{k_1\to0}{\partial\over\partial k_1^i}D_{k_1}D_{k_2}D_{k_3} W_0&=-k_2^a{\partial^2\over\partial k_2^a\partial k_2^i}D_{k_2}D_{-k_2} W_0+\frac{1}{2}k_2^i{\partial^2\over\partial k_2^a\partial k_2^a}D_{k_2}D_{-k_2} W_0=\cr&=-k_2^i\left({1\over k_2}{\partial\over\partial k_2} -{1\over2}{\partial^2\over\partial k_2^2}\right)D_{k_2}D_{-k_2} W_0
}}
Where we used that the second derivative of the wavefunction depends only on the absolute value of $k_2$. Then, plugging this back in \expfirst\ we get:
\eqn\first{
F^a(k_2)=-{1\over2}\partial_{k_2^a}Z(k_2)={1\over2}\partial_{k_2^a}\left[\langle\zeta_{k_2}\zeta_{-k_2}\rangle{\partial \log \left[k_2^3  \langle \zeta_{k_2}\zeta_{-k_2}\rangle\right] \over \partial \log k_2}\right]
}
It was observed in \CreminelliRH\ that under the substitution $k_1\to k_L$, $k_2 \to k_S-k_L/2$, the linear term in \tayexp , $F^a(k_S)$ is absent. One can check that changing variables from $k_2$ to $k_S$ such is the case, so \first\ is compatible with the claims made in \CreminelliED\CreminelliRH .

We can also study the case of one long tensor mode and two short scalar modes, as in \CreminelliED\foot{In particular, we want to check equations (66) and (67) of \CreminelliED .}. There is a small point to be made, which is the following. We obtain from our method the object $\partial_{k^a}D_k^{bc}DDW_0$. Then, we need to contract this with a polarization tensor. But the expectation value we consider is already contracted with the polarization tensor, so we could be neglecting a term where the derivative acts on the polarization tensor, and the resulting tensor is contracted with the expectation value. In other words, we do not calculate the contribution coming from $(\partial_{k^a}\epsilon^{bc})D_k^{bc}DDW_0$. We show in appendix C that this contribution is zero, and so we capture the entire linear term in the long momentum. The consistency condition to linear order will be:
\eqn\firsttens{
\lim_{k_1\to0}\langle \gamma^s_{k_1} \zeta_{k_2}\zeta_{k_3}\rangle=\langle\gamma^s_{k_1}\gamma^s_{k_1}\rangle\left\{Z_\gamma(k_2)+{k_1.k_2\over 4k_2^2}{k_2.\epsilon_1.k_2\over k_2^2} \left[k_2\partial_{k_2}-k_2^2\partial^2_{k_2}\right]\langle\zeta_{k_2}\zeta_{k_2} \rangle\right\}
}
With $Z_\gamma(k_2)$ can be read out from \tpfmixlt . This result agrees with the prescription given in \CreminelliED , and, in particular, with the case of single field inflation\MaldacenaVR . 

To obtain the higher order consistency conditions described in \HinterbichlerDPA , note the following. It is clear that, taking multiple derivatives with respect to the momentum being squeezed, the best we can do is obtain an expression for the symmetrized derivative $\partial^{(i_1}\partial^{i_2}\cdots\partial^{i_{n-1}} D^{i_n)~j}_{k_1}D_{k_2}D_{k_3}W_0$. We can project out some components of this symmetrized derivative, and relate it to linear combinations of three point functions, as is done in \HinterbichlerDPA . Of course, one would not expect to be able to obtain all derivatives of the wavefunction from the Ward identity, as we are just finding the parts of the expectation value fixed by gauge invariance.

Let us study in more detail the case of the second derivative. We obtain:
\eqn\quadconall{
\lim_{k_1\to0}\left[{\partial^2\over\partial k_1^a\partial k_1^b}D^{cj}_{k_1}+{\partial^2\over\partial k_1^c\partial k_1^a}D^{bj}_{k_1}+{\partial^2\over\partial k_1^b\partial k_1^c}D^{aj}_{k_1}\right]D_{k_2}D_{k_3} W_0=-{1\over2}k_2^j{\partial^3\over\partial k_2^a\partial k_2^b\partial k_2^c}
D_{k_2}D_{-k_2} W_0}

There are two types of indices in some sense here, the index that is not symmetrized and the symmetrized ones. Just as we did for the first derivative, we can either contract symmetrized indices or one symmetrized index with the separate one. Take $\delta^{ab}$ and $\delta^{ac}$, and after some manipulation, the best one can obtain is (we write explicitly $D^{cc}$, so there is no confusion with derivatives with respect to the scalar, $D\equiv 2 D^{cc}$):

\eqn\quadtent{\eqalign{
\lim_{k_1\to0}&\left[{\partial^2\over\partial k_1^a\partial k_1^b}D^{cc}_{k_1}-{\partial^2\over\partial k_1^c\partial k_1^c}D^{ab}_{k_1}\right]D_{k_2}D_{k_3} W_0=\cr=&-{1\over2}\left[k_2^c{\partial^3\over\partial k_2^c\partial k_2^a\partial k_2^b}-k_2^a{\partial^3\over\partial k_2^b\partial k_2^c\partial k_2^c}\right]
D_{k_2}D_{-k_2} W_0}}

Which is not good enough to isolate what we would like, $\lim_{k_1\to0}{\partial^2\over\partial k_1^a\partial k_1^b}D_{k_1}D_{k_2}D_{k_3} W_0$.

\newsec{Comment on gauge/gravity duality}

In gauge/gravity duality, $dS/CFT$ \refs{\WittenKN , \StromingerPN , \MaldacenaVR} is the proposal that an asymptotically de Sitter space can be described by a dual field theory. In the approach of \MaldacenaVR , the proposal  is that the wavefunction of the universe with a certain 3-metric profile is equal to the partition function of a CFT, where this 3-metric is a parameter of the partition function.

Then, the stress tensor of the dual field theory, $T^{ij}$, is given by functional derivatives of the partition function with respect to the metric. So, when we take functional derivatives of $\Psi$ with respect to the flat background, in the dual picture we are computing correlation functions of the stress tensor in the vacuum of the field theory. So, from the field theory perspective, the functional derivatives of the metric we considered throughout the paper, $D^{ij}D^{kl}\cdots \Psi$ are equal to correlation functions of the stress tensor, $\langle T^{ij}T^{kl}\cdots\rangle$. 

From this point of view, the consistency condition has a simple interpretation. \wig\ expresses the conservation of the stress tensor of the dual theory, $\nabla_i\langle T^{ij}\rangle=0$. So, \wig\ is equivalent to Ward identities obeyed by the stress tensor, which can be found in \OsbornCR . Note, though, that we do NOT need anything like dS/CFT or gauge/gravity duality to use the Ward identities that the $T^{ij}$ satisfy. These have a pure bulk interpretation from diffeomorphism invariance. 

Note also that the final equations in \wloc\ and \wlocsc\ can be interpreted as identities obeyed by the trace of the stress tensor. \wloc\ states that an insertion of the trace of the stress tensor should render any correlation function to be zero. This means that $\langle T^{ii} \cdots \rangle=0$. This is expected, as de Sitter has isometries at late times that are isomorphic to the conformal group \MaldacenaNZ . For the single field case, there is no conformal symmetry, due to the presence of the inflaton. It corresponds to the insertion of an operator that deforms the CFT \refs{\MaldacenaVR , \LarsenET , \LarsenPF , \LarsenKF}. This operator breaks the conformal symmetry, and induces a trace to the stress tensor. The relation between the operator and the trace is given by the last equation of \wlocsc .
\newsec{Discussion}

In this paper, we gave a different perspective on how to derive inflationary consistency conditions. The objectives of this approach were two-fold. First, to show that the origin of these conditions stems from diffeomorphism invariance of the wavefunction. Second, this approach seems to be generalizable to other inflationary theories, and thus could be exploited in more generality, in the same fashion that Ward identities are derived from symmetries of the path integral.

It is important to notice that we are always dealing with the ``mathematical" squeezed limit, in the sense of taking the long mode wave number to zero. There are several models where the consistency condition is violated, in the sense of the ratio of the sides of the triangle being small, but not zero. This physical squeezed limit can probe different scales in the theory, and is usually associated to the long modes not freezing at this scale. It would be interesting to use the methods in \CreminelliCGA\ to see if one can say something in general about the leading order term in theories that violate the consistency condition. Let us also observe that, throughout the paper, we used a technical assumption, namely, that $\lim_{k\to0} {d\over d \log k} D_k D\cdots D\Psi = 0~$\foot{This assumption seems to have no drawbacks for the following reason: a term that violates it would have to be an analytic function of one of the soft momenta. Thus, it is a contact term in position space and is not the piece fixed by these consistency conditions. Also, note that if we have a field theory that produces an almost scale invariant spectrum, either the logarithm is the indicator of an anomalous dimension coming from loop effects, or it should be discarded as it comes from a contact term in the expectation value. Note that a term of the form $\log(k_1+k_2+k_3)$ is allowed on a three point function, as it satisfies the assumption we made. An argument based on analyticity was made in \BerezhianiEWA\ that such terms would violate the attractor structure of single field models.}.

The consistency condition can be also stated in terms of modes that are still inside the horizon. In our language we are always dealing with the superhorizon wavefunction. In a semiclassical setup, where cosmology is treated as an effective field theory, we can evaluate the semiclassical wave function at a given time slice, $\Psi[h,\phi,\eta]$. It is not clear that this object makes sense beyond effective field theory, but for the purposes of studying inflation as an effective theory, it is well defined and one can follow the same steps of the previous sections. The long mode would still correspond to taking $k\to0$, but the other modes in the expectation value will be inside the horizon and the same consistency conditions would follow. The novelty here is that expectation values with derivatives in the subhorizon modes are non-zero, and thus one can derive new consistency conditions for those. Still, they should follow from the same diffeomorphism invariance constraint.
 
The leading terms of these inflationary expectation values are thus fixed by gauge invariance. Of course, it would be nice if we could compute observables free of these pure gauge pieces. In gauge theory we know the answer to this question. In gravitational theories the answer is not so clear, unless there is a dual description in terms of a field theory. From the wavefunctional point of view, this would be equivalent to regarding its derivatives as the fundamental observables. In gauge gravity duality, these would translate to expectation values of the stress tensor of the theory. There, the consistency condition has the interpretation that, at zero momentum, there is an ambiguity related to the definition of the stress tensor \RouetUT .

The analysis carried in this note involved tree level expectation values. But the general structure of the Ward Identities coming from coordinate reparametrization invariance are valid to any loop order. The Hamiltonian constraint would necessarily involve some UV regularization, which leads to renormalization of observables etc. (this has been already carried out for gauge theories in \MansfieldPD ). But for the one-loop cosmology and beyond, the momentum constraint remain unchanged. So there might be some generalized consistency condition to n-loops. Maybe it can be generated recursively, just like in the recent proofs of conservation of the inflaton $\zeta$ outside of the horizon, which rely on the consistency condition \refs{\SenatoreCF , \SenatoreNQ , \PimentelTW , \SenatoreYA}.
\bigskip
{\bf Acknowledgments:} It is a pleasure to thank J. Maldacena, E. Pajer and M. Zaldarriaga for very useful discussions. I also thank J. Maldacena and M. Zaldarriaga for comments on a draft of this paper. G.L.P. was supported by the Department of State through a Fulbright Science and Technology Fellowship and through the U.S. NSF under Grant No. PHY-0756966. 
 
\appendix{A}{Four point functions and beyond}

The procedure in the paper can be generalized to higher order expectation values. Here we outline the general features of this procedure. The main difference is that there are two different squeezed limits. Namely, the limit when an external leg has zero momentum, or when an internal leg has zero momentum - a collinear limit. 

Let us illustrate that point by considering a four point expectation value of scalar fluctuations in single field inflation. Its form, in terms of derivatives of the wavefunction, is given by\foot{We consider only the connected part of the four point expectation value. There is an additional contribution coming from disconnected diagrams, which is present in the free theory, which is thus Gaussian.}:
\eqn\fpexp{\eqalign{
\langle \zeta_{k_1}\zeta_{k_2}\zeta_{k_3}\zeta_{k_4}\rangle &= \frac{1}{\Pi_i 2{\rm Re} \left[D_{k_i}D_{-k_i}W_0\right]}\left\{2 {\rm Re}[D_{k_1}D_{k_2}D_{k_3}D_{k_4}W_0] +\right.\cr&\left.+\sum_a \frac{2 {\rm Re}[D_{k_1}D_{k_2}D^a_{-k_1-k_2}W_0]2 {\rm Re}[D^a_{-k_3-k_4}D_{k_3}D_{k_4}W_0]}{ 2{\rm Re}\left[D^a_{k_1+k_2}D^a_{-k_1-k_2}W_0\right]}+{\rm permutations}\right\}
} }
Where the $\sum_a$ represents the sum over all degrees of freedom (two graviton polarizations and one scalar). The permutations account for the different exchange channels for the internal leg, like the $s$, $t$ and $u$ channels in four particle scattering. 

The external squeezed limit is completely analogous to the one in the main text. One needs to commute three functional derivatives through the momentum constraint and take the limit of an external momentum to zero etc. We analyze the internal momentum squeezed limit in detail, as it has no analogue for the three point expectation value. 

Consider the limit $k_1 \to -k_2$. Of course, due to translation invariance, $k_4 \to -k_3$. We see that the overall denominator in \fpexp\ does not diverge. The only singular piece comes from the exchange diagrams that involve two vertices, as its denominator involves a second derivative of $W_0$ evaluated at $k_1+k_2$ momentum. The contributions from the four-derivative term and other exchange terms are thus subleading. For the leading term, its numerator is the square of the squeezed limit of a third derivative, so we use \conallfo\ to relate these to two point expectation values. Thus, to leading order in $k_1+k_2$, we obtain:
\eqn\fpcoll{\eqalign{
&\lim_{k_2\to-k_1}\langle \zeta_{k_1}\zeta_{k_2}\zeta_{k_3}\zeta_{k_4}\rangle =\cr&= \langle \zeta_{k_1+k_2}\zeta_{k_1+k_2} \rangle \left(\langle\zeta_{k_1}\zeta_{-k_1}\rangle{\partial \log \left[k_1^3  \langle \zeta_{k_1}\zeta_{-k_1}\rangle\right] \over \partial \log k_1}\right)\left(\langle\zeta_{k_3}\zeta_{-k_3}\rangle{\partial \log \left[k_3^3  \langle \zeta_{k_3}\zeta_{-k_3}\rangle\right] \over \partial \log k_3}\right)+\cr&+\sum_s\langle\gamma^s_{k_1+k_2}\gamma^s_{k_1+k_2} \rangle \left(\epsilon^{s,~ij}_{k_1+k_2}k_1^ik_1^j {\partial \over \partial k_1^2}\langle \zeta_{k_1}\zeta_{-k_1}\rangle\right)\left(\epsilon^{s,~ij}_{k_1+k_2}k_3^ik_3^j {\partial \over \partial k_3^2}\langle \zeta_{k_3}\zeta_{-k_3}\rangle \right)
}}
Which is consistent with the results described in \refs{\SeeryAX, \LeblondYQ, \SenatoreWY}.

Note that this procedure can be extended to an n-point expectation value, but the amount of diagrams contributing beyond leading order makes the general expressions become cumbersome. This problem is treated in detail in \HinterbichlerDPA . There, a prescription to calculate the contribution from diagrams with exchanged particles, like the one we consider for the collinear limit, is given in detail.

\appendix{B}{Massless scalar spectator field}
Let us analyze an example of an inflationary theory with an inflaton plus a massless scalar field, similar to the example discussed in \WeinbergVY . We have the metric, $h_{ij}$, the inflaton $\phi$ and the spectator field $\sigma$. The condition of reparametrization invariance of the wavefunction is a simple extension of \WdWsctr :
\eqn\WdWmul{
2 \nabla_i\left[\frac{1}{\sqrt{h}}\frac{\delta W_0}{\delta h_{ij}} \right]-{1\over\sqrt h}h^{ij}\partial_i\phi\frac{\delta W_0}{\delta \phi}-{1\over\sqrt h}h^{ij}\partial_i\sigma\frac{\delta W_0}{\delta \sigma}=0
}

As in the single field case, we are interested on the wavefunction calculated on a slice of constant inflaton field. So we are effectively treating a momentum constraint of the form:
\eqn\WdWmulef{
2 \nabla_i\left[\frac{1}{\sqrt{h}}\frac{\delta W[h(x),\phi,\sigma(x)]}{\delta h_{ij}} \right]-{1\over\sqrt h}h^{ij}\partial_i\sigma\frac{\delta W[h(x),\phi,\sigma(x)]}{\delta \sigma}=0
}

Now, we take two functional derivatives with respect to the massless field, say, $\delta^2/\delta\sigma(y)\delta\sigma(z)$. They commute through the covariant derivative, and each one may hit the $\partial_i\sigma(x)$ term in \WdWmulef . Because we are dealing with a scalar operator, it should be no surprise that we obtain a Ward identity identical to \scgen . After going to momentum space, and using the $D$ notation for the functional derivatives, we finally obtain:
\eqn\scphigen{
k_1^aD^{aj}_{k_1}D^\sigma_{k_2}D^\sigma_{k_3} W_0={1\over2}\left[-k_2^j D^\sigma_{k_3}D^\sigma_{-k_3} W_0-k_3^jD^\sigma_{k_2}D^\sigma_{-k_2} W_0\right]
}
So we have the same consistency condition as the one for scalar operators, namely:
\eqn\consphi{
\lim_{k_1\to0}\langle \zeta_{k_1}\sigma_{k_2}\sigma_{k_3}\rangle = \langle\zeta_{k_1}\zeta_{-k_1}\rangle {\partial \log \left[k_2^3  \langle \sigma_{k_2}\sigma_{-k_2}\rangle\right] \over \partial \log k_2}=-n_\sigma  \langle\zeta_{k_1}\zeta_{-k_1}\rangle \langle \sigma_{k_2}\sigma_{-k_2}\rangle
}
The field $\sigma$ behaves as a free field in (quasi) de Sitter space, so its spectral index is the same as that of a gravitational wave, given by $n_\sigma=-2\epsilon$ \MaldacenaVR , where the slow roll factor is related to the variation of Hubble's constant, $\epsilon H^2= - \dot H$. In fact, the computation of the three point function and the check for the squeezed limit is quite similar to the case of two gravitons and one scalar studied in \MaldacenaVR . 

To check \consphi\ we need to compute the three point expectation value using the in-in formalism. We just state the main equations here. The quadratic actions, with corresponding (late time) two point functions are given by:
\eqn\quadzephi{\eqalign{
&S_{\zeta\zeta}={1\over2}\int dt d^3x (2\epsilon)\left[a^3\dot\zeta^2-a(\partial\zeta)^2\right],~~~~~~\langle\zeta_{k_1}\zeta_{k_2}\rangle=(2\pi)^3\delta^3(k_1+k_2) \frac{H^2}{4\epsilon k^3}\cr
&S_{\sigma\sigma}={1\over2}\int dt d^3x  \left[a^3\dot\sigma^2-a(\partial\sigma)^2\right],~~~~~~~~~~\langle\sigma_{k_1}\sigma_{k_2}\rangle=(2\pi)^3\delta^3(k_1+k_2) \frac{H^2}{2 k^3}
}}

 The cubic action is given by equation (27) of \WeinbergVY . We can integrate it by parts, following the strategy in \MaldacenaVR , to see that the interaction is of order $\epsilon$, the slow roll parameter:
\eqn\weinbint{\eqalign{
S_{\zeta\sigma\sigma}=&\int\left[-{a\over2}\zeta(\partial \sigma)^2-{a\over2H}\dot\zeta(\partial\sigma)^2+a\partial_i\left({\zeta\over H}-\epsilon a^2 \partial^{-2}\dot\zeta\right)\dot\sigma\partial_i\sigma-\right.\cr&\left.-{a^3\over 2H} \dot\zeta\dot\sigma^2+{3 a^3\over 2}\zeta\dot\sigma^2\right]=\int\left\{\epsilon\left[-\zeta L_{\sigma\sigma} - a^3 \partial_i \partial^{-2}\dot\zeta\dot\sigma\partial_i\sigma\right]+{\zeta\dot\sigma\over H}{\delta L_{\sigma\sigma}\over\delta\sigma}\right\}
}}
The last term is proportional to the quadratic equations of motion, and can be removed from the action by a proper $\sigma$ field redefinition\MaldacenaVR . This redefinition does not alter the final three point expectation value though, as it vanishes outside the horizon. What is left are the terms in square brackets. It is clear that, when the $\zeta$ mode becomes very long in wavelength, it is simply rescaling the two point Lagrangian for the $\sigma$ field. This is the standard bulk intuition to justify the consistency relation. In fact, one can check that the second term in square brackets is subleading in the squeezed limit, thus making this intuition rigorous. To leading order in slow roll, the squeezed three point expectation value is given by:
\eqn\checklight{
\lim_{k_1\to0}\langle\zeta_{k_1}\sigma_{k_2}\sigma_{k_3}\rangle={H^4\over 4 k_1^3 k_2^3}=-(-2\epsilon){H^2\over 4\epsilon k_1^3}{H^2\over 2 k_2^3}= -n_\sigma  \langle\zeta_{k_1}\zeta_{-k_1}\rangle \langle \sigma_{k_2}\sigma_{-k_2}\rangle
}
An important observation is the following. Because the massless $\sigma$ field is free to fluctuate, it can convert itself into $\zeta$ fluctuations during reheating and other phases beyond inflation. Also, we assumed that there is a quasi-de Sitter background over which the fluctuations evolve. Thus, it is not necessarily true that the three point expectation values computed here are kept frozen and will induce temperature correlations in the CMB.

\appendix{C}{Derivative of the polarization tensor}
Take a three point expectation value that involves a long tensor mode. If we expand it to linear order in the long mode, we obtain:
\eqn\explm{
\langle \gamma^s_{k_1}\zeta_{k_2}\zeta_{k_3}\rangle=\langle \gamma^s_{0}\zeta_{k_2}\zeta_{-k_2}\rangle+k_1^a\left[\langle\gamma^{bc}_{k_1}\zeta_{k_2}\zeta_{k_3}\rangle{\partial\over\partial k_1^a}\epsilon_{bc}^1+\epsilon_{bc}^1{\partial\over\partial k_1^a}\langle\gamma^{bc}_{k_1}\zeta_{k_2}\zeta_{k_3}\rangle\right]+\cdots
}
We want to show that the first term does not contribute in brackets does not contribute to \explm . In order to do that, we take derivatives of the defining expressions for the polarization tensor \poltens\ and obtain:
\eqn\pold{
\left({\partial\over\partial k^a}\epsilon_{bc}\right)\epsilon_{bc}=0,~~~\epsilon_{ab}+k_c{\partial\over\partial k^a}\epsilon_{bc}=0 \Rightarrow {\partial\over\partial k^a}\epsilon_{bc}=-{k_b\epsilon_{ac}+k_c\epsilon_{ab}\over k^2}
}
Which, when contracted with $k_1^a$, will give zero. This is why we are free to contract the polarization tensor directly with ${\partial\over\partial k_1^a}\langle\gamma^{bc}_{k_1}\zeta_{k_2}\zeta_{k_3}\rangle$ to derive the linear consistency condition.

\listrefs
\bye